\documentclass[amsmath,amssymb,aps,prl,floatfix,twocolumn,showpacs,superscriptaddress]{revtex4}

\usepackage{amsmath}
\usepackage{amssymb}
\usepackage{graphicx}
\usepackage{bm}
\usepackage{float}
\usepackage{calc}
\usepackage{subfig}

\newcommand{\diff}{\text{d}}
\newcommand{\mean}[1]{\langle #1 \rangle}
\newcommand{\meqref}[1]{{Eq.~(\ref{#1})}}
\newcommand{\Gone}{{{$\mathcal G1$}}}
\newcommand{\Gtwo}{{{$\mathcal G2$}}}
\newcommand{\Gthree}{{{$\mathcal G3$}}}

\begin{document}

\title{Role of anisotropy for protein-protein encounter}
\author{Jakob Schluttig}
\affiliation{University of Heidelberg, Institut f\"ur theoretische Physik, Philosophenweg 19, 69120 Heidelberg, Germany}
\affiliation{University of Heidelberg, Bioquant, Im Neuenheimer Feld 267, 69120 Heidelberg, Germany}
\author{Christian Korn}
\affiliation{University of Heidelberg, Bioquant, Im Neuenheimer Feld 267, 69120 Heidelberg, Germany}
\author{Ulrich S. Schwarz}
\affiliation{University of Heidelberg, Institut f\"ur theoretische Physik, Philosophenweg 19, 69120 Heidelberg, Germany}
\affiliation{University of Heidelberg, Bioquant, Im Neuenheimer Feld 267, 69120 Heidelberg, Germany}

\begin{abstract}
Protein-protein interactions comprise both transport and reaction
steps. During the transport step, anisotropy of proteins and their
complexes is important both for hydrodynamic diffusion and
accessibility of the binding site.  Using a Brownian dynamics approach
and extensive computer simulations, we quantify the effect of
anisotropy on the encounter rate of ellipsoidal particles covered with
spherical encounter patches. We show that the encounter rate $k$
depends on the aspect ratios $\xi$ mainly through steric effects,
while anisotropic diffusion has only a little effect. Calculating
analytically the crossover times from anisotropic to isotropic
diffusion in three dimensions, we find that they are much smaller than
typical protein encounter times, in agreement with our numerical
results.
\end{abstract}


\maketitle

Protein-protein interactions are at the heart of most molecular
processes in biological systems and are intensively investigated both
by experiment and theory \cite{schreiber_09}. Conceptually,
protein-protein binding consists of transport and reaction
steps. Brownian dynamics simulations have been introduced to study
transport towards association of spherical model particles
\cite{ermak_78,northrup_92}. Isotropy of the diffusion process can be
used to develop computer time efficient methods for propagating
reactive particles in time and space \cite{wolde_05_a}. However,
assuming spherical particles does not address two important aspects of
protein encounter. First, proteins and their complexes are typically
not spherical, but their shape and diffusional behaviour can be highly
anisotropic. Second, their binding sites are strongly localized and
thus binding is not isotropic neither. Both of these issues have been
addressed in Brownian dynamics simulations, e.g.\ by using approximate
schemes for the mobility tensor of arbitrarily shaped proteins
\cite{torre_00} and by incorporating all-atom force fields
\cite{gabdoulline_97}.  However, the full role of anisotropy for
protein encounter has not been systematically studied before in a
generic model for protein-protein encounter. Although not suited to
answer specific biological questions, simple models without atomic
details also have the advantage that they can be easily upscaled to
large system sizes.

\begin{figure}[t]
  \begin{center}
  \includegraphics[width=0.475\textwidth]{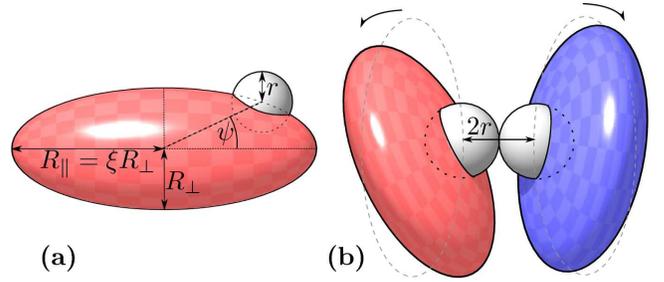}
  \end{center}
  \caption{(a) Geometry of an ellipsoidal model particle with a
    reactive patch on its surface at some angle $\psi$ with respect to
    the symmetry axis $\mathbf e_\parallel$. (b) Illustration of an
    encounter configuration.
    \label{ellipsoid_basic_sketch}
  }
\end{figure}

In order to study theoretically the effect of shape anisotropy on
protein-protein encounter, we systematically vary the aspect ratio
$\xi=R_\parallel/R_\perp$ of ellipsoids with a rotational symmetry,
compare Fig.\ \ref{ellipsoid_basic_sketch}(a). This is one of the few
systems for which closed analytic expressions of the translational
($^t$) and rotational ($^r$) friction coefficients
$\zeta^{t/r}_{\parallel/\perp}(R_\parallel,R_\perp)$ are known
\cite{perrin_34, perrin_36}. Recently, the diffusional properties of
such anisotropic particles have been investigated in great detail for
two dimensions \cite{han_06,munk_09}. Here we study three dimensions,
both numerically and analytically. In addition, we not only consider
anisotropy in diffusion, but also anisotropy in binding by placing
spherical encounter patches at various positions on the ellipsoids.
This approach allows us to assess in quantitative detail the relative
role of hydrodynamic and steric anisotropy for protein-protein
encounter.


\begin{figure*}[t]
  \begin{center}
  \includegraphics[width=0.94\textwidth]{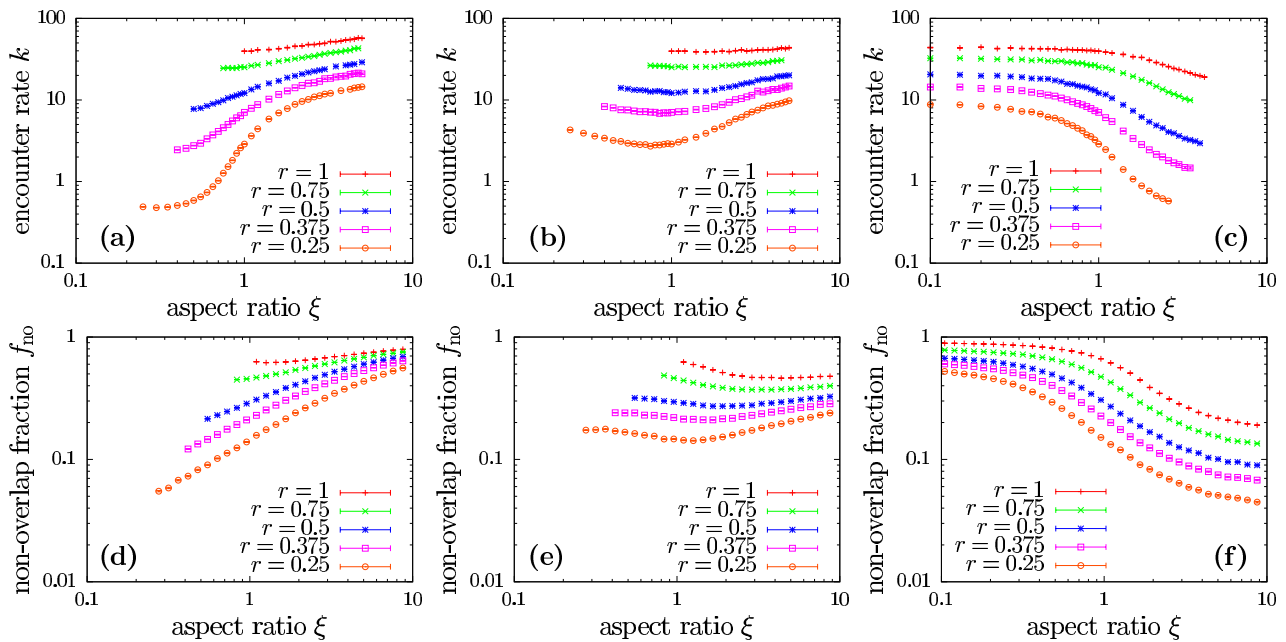}
  \end{center}
  \caption{ \label{ellipsoid_data_plot} (a)--(c) Encounter rates
    for different patch sizes and locations. (d)--(f) Statistical estimate of the fraction of
    configuration space available for encounter.  The patches are
    located according to \Gone\ in (a) and (d), \Gtwo\ in (b) and (e) and
    \Gthree\ in (c) and (f). }
\end{figure*}

To quantify the effect of anisotropy on molecular encounter, we
performed Brownian dynamics simulations with a pair of ellipsoids in a
periodic boundary box of edge length $L$, which represents the
concentration of our system. The appropriate Langevin equation is
evolved in discrete time steps $\Delta t$ via the Euler algorithm
\cite{schluttig_08}:
\begin{align}
\mathbf X(t+\Delta t)=\mathbf X(t)+\mathbf g(\Delta t)+\mathcal
O(\Delta t^2) \text{ .}
\end{align}
All vectors are six-dimensional generalized coordinates including the
orientational information. The noise term $\mathbf g(\Delta t)$ is
determined by the Einstein relation $\langle\mathbf g\rangle=\mathbf
0$, $\langle \mathbf g(\Delta t)\cdot\mathbf g^\dagger(\Delta
t)\rangle=2k_BT_a\mathbb M \Delta t$, where $T_a$ is ambient
temperature and $\mathbb M=\mathbb \zeta^{-1}$ is the $6\times6$
mobility matrix. In the following, all lengths are given in units of
$U_L=R_\perp$ and times are scaled by
$U_T=\zeta^r(R_\perp,R_\perp)/(k_BT_a)$. In the simulations we chose
$U_L=2\mathrm{nm}$ and thus $U_T\approx 50 \mathrm{ns}$ at viscosity
of water $\eta=1\mathrm{mPa\,s}$. As time step we use $\Delta
t=2\cdot10^{-5}$. We do not consider direct forces between the model
particles in this study; in particular, we neglect electrostatic and
two-body hydrodynamic interactions. Otherwise we implement hard-core
repulsion.  An analytic overlap criterion for a pair of ellipsoids is
difficult to derive, but suitable algorithms based on the solution of
the characteristic equation of the two ellipsoids have been derived
for hard body fluids \cite{allen_93}.

Each of the model particles carries a spherical reactive patch of
radius $r$ on its surface whose location is described by the angle
$\psi$, compare Fig.\ \ref{ellipsoid_basic_sketch}(a). An overlap of
these reactive patches is considered as an encounter, compare
Fig.\ \ref{ellipsoid_basic_sketch}(b). We average over all initial
conditions by starting $10^4$ simulations at random initial positions
and orientations for each parameter set. The main quantity of interest
is the encounter rate $k$, defined as the inverse first passage time
to encounter. As it is common in Brownian dynamics of proteins, the
encounter rate thus emerges from the definition of an absorbing
boundary for the random walk \cite{schreiber_09}. For the two model
particles in the simulation box we consider three scenarios of patch
locations: $\psi_1=\psi_2=0$ \label{abbr_g1g3} (\Gone); $\psi_1=0$,
$\psi_2=\pi/2$ (\Gtwo); $\psi_1=\psi_2=\pi/2$ (\Gthree).

In a coordinate system spanned by the principal axes, the friction
matrix of an ellipsoid is diagonal. Taking the corresponding values
for a sphere of radius $R_\perp$ as a reference scale, the diffusion
coefficients $D^{t/r}_{\parallel/\perp}$ only depend on $\xi$. The
relative translational mobility of an ellipsoid compared to a sphere
is then given as $\lambda(\xi)=D^t_\parallel(\xi)+2D^t_\perp(\xi)$.
According to the Smoluchowski equation, the encounter rate is expected
to depend linearly on the mobility and the concentration of the
reacting particles. Therefore we normalize the encounter rates
obtained by our simulations by $\lambda(\xi)$ and by the volume of our
simulation box (here $L^3=100^3$). The corresponding data is shown in
Figs.\ \ref{ellipsoid_data_plot}(a)--(c). The rates are given in
$1/U_T$. In each case, we exclude aspect ratios for which the reaction
patches span the whole ellipsoid. Our simulation results show that
encounter rates can vary up to two orders of magnitude depending on
aspect ratio and patch position. While for patches located at the tips
(\Gone), encounter efficiency increases with aspect ratio, it decreases
for patches located at the sides (\Gthree). For the mixed case
(\Gtwo), changes in aspect ratio have only a weak effect.

When interpreting these results, an important systematic
difference between different aspect ratios $\xi$ has to be noted: As
the geometry of the ellipsoid is changing with $\xi$, the exposed
volume fraction \label{symb_patchaccess} $f_V$ of the reactive patches
not covered by the steric particle is changing. We estimated the
steric effect on the encounter rate by studying the fraction of
non-overlapping ellipsoid configurations with touching reactive
patches $f_\mathrm{no}$. A scheme of the setup is shown in
Fig.~\ref{ellipsoid_basic_sketch}(b). Particularly, the centers of the
patches are placed at the distance $2r$ and both ellipsoids are
randomly rotated around the center of their respective patch. The
results for $f_\mathrm{no}$ from drawing $10^5$ of such random
encounter configurations are shown in
Figs.~\ref{ellipsoid_data_plot}(d)--(f) for the parameters
corresponding to Figs.~\ref{ellipsoid_data_plot}(a)--(c). The plots
show that the qualitative features of the encounter rates are well
reproduced by the steric constraints.  This leads to the conclusion
that the main reason for the changes in the encounter rate in the
preceeding study is not the altered hydrodynamic behavior of the
ellipsoids but the steric hindering of encounters due to the changing
geometry.


\begin{figure*}[t]
  \begin{center}
  \includegraphics[width=0.94\textwidth]{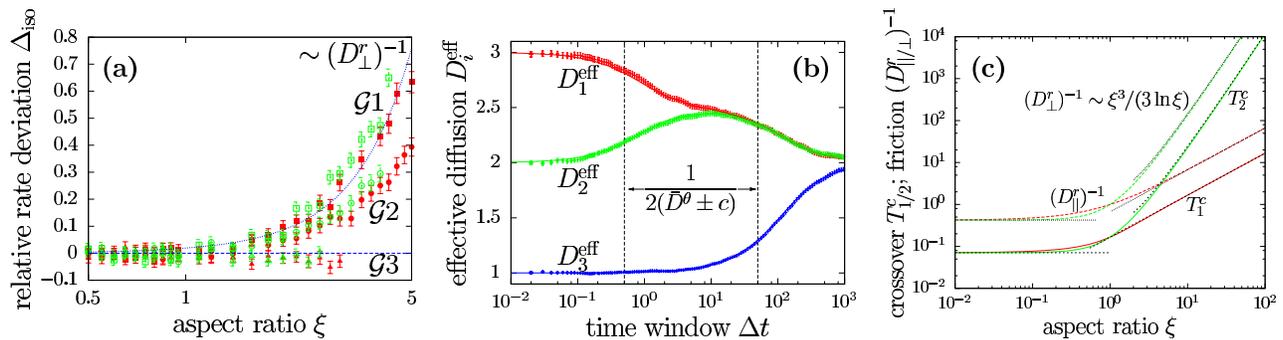}
  \end{center}
  \caption{
    \label{aniso_iso_compare}
    (a) Relative deviation of the encounter rates from the original data (compare
    Fig.~\ref{ellipsoid_data_plot}) assuming
    isotropic motion at all times. Reactive patches have a radius of
    $r=0.5$. The dotted line indicates the supposed scaling and the
    dashed line marks $\Delta_\mathrm{iso}=0$. Data from simulations with $L=100$
    is shown with red, filled symbols, data with $L=20$ is shown with
    green, hollow symbols. (b) Example of the time window
    dependency of the effective principal translational diffusion
    coefficients for the following choice of parameters (generic
    units): $D_1=3$, $D_2=2$, $D_3=1$, $D^\theta_1=0$,
    $D^\theta_2=0.005$, $D^\theta_3=0.5$. The data points have been
    obtained by simulation (the error bars depict the standard
    deviation obtained from $10^5$ individual runs), the solid lines
    represent the theoretic prediction and the dashed lines indicate
    the two relevant time scales for the crossover as given in
    \meqref{crossover_finalgdef}. (c) Crossover times and
    rotational friction coefficients for ellipsoids for a wide range of
    aspect ratios.}
\end{figure*}

To obtain another measure for the effect of hydrodynamics, we
compared our simulations with explicit anisotropic diffusion to
simulations, where we do \emph{not} account for the anisotropy in the
diffusion matrix. Particularly, we use an isotropic diffusion matrix
with the average translational and rotational diffusion coefficients
$D^t_\mathrm{avg}=(D^t_\parallel+2D^t_\perp)/3$, which is equal to the
isotropic limit, and $D^r_\mathrm{avg}=(D^r_\parallel+2D^r_\perp)/3$,
respectively. We perform the same simulations as in
Fig.\ \ref{ellipsoid_data_plot} with this new $\mathbb D$. Moreover,
we compared to simulations at higher effective concentration
($L=20$). Fig.\ \ref{aniso_iso_compare} shows the relative deviation
of the encounter rates
$\Delta_\mathrm{iso}=k_\mathrm{isotropic}/k_\mathrm{anisotropic}-1$. Interestingly,
in case \Gthree\ there is no significant deviation from the original
results. However, considering \Gone\ the (artificial) isotropic
encounter rate is larger for prolates ($\xi>1$). This effect can also
be observed in the mixed case \Gtwo, decreased by roughly a factor
$2$. This is reasonable as here only one of the two encountering
ellipsoids has its patch at $\psi=0$. These findings show that the
anisotropic diffusion of elongated ellipsoids leads to a decrease of
the encounter rates. However, the deviation up to $\xi=5$ is
moderate ($\Delta_\mathrm{iso}\approx0.7$), so we conclude
that the impact of anisotropic diffusion on molecular encounter is
rather weak. 


Anisotropic diffusion in isotropic environments is only relevant on
small time and length scales. In the following, we derive the
effective translational diffusion properties for a finite time window
$\Delta t$ considering an arbitrary body with three different
translational ($D_i$) and rotational ($D^\theta_i$) diffusion
coefficients along the principal axis (in the body-fixed coordinate
system). As the rotations due to rotational diffusion are supposed to
be completely independent, the crossover only affects the effective
translational diffusion which is described by a $3\times3$ diffusion
matrix: $\mathbb{D}^t_{ij}=\delta_{ij}D_i$, where $\delta_{ij}$ is the
Kronecker delta. A rotation determined by a vector of angles $\bm
\theta=(\theta_1,\theta_2,\theta_3)^\dagger$ around the three
principal axes can be described by the rotation matrix
$\mathbb{S}=e^{-\bm\theta\cdot\mathbf{J}}$, with $ \mathbf{J}= \left(
\mathbb{J}^1 \text{, } \mathbb{J}^2 \text{, } \mathbb{J}^3
\right)^\dagger$.  $\bm\theta\cdot\mathbf{J}$ denotes a formal scalar
product and $\mathbb{J}^k$ are matrices defined by
$\mathbb{J}^k_{ij}=\epsilon_{ikj}$, where $\epsilon$ is the
Levi-Civita symbol. We proceed considering only small rotations
occurring at small times, so that we can expand the rotation in orders
of $\theta$:
\begin{align}
\mathbb{S}_{ij}=\delta_{ij}\frac{2-\bm\theta^2}{2}
+\epsilon_{ikj}\theta_k\frac{6-\bm\theta^2}{6}
+\frac{\theta_i\theta_j}{2}+\mathcal O(\theta^4)
\text{ ,}
\label{needle_rotation_matrix}
\end{align}
where $\bm\theta^2=\sum_{i=1}^3\theta_i^2$.  This rotation is applied
to $\mathbb{D}^t$ and we get
$\bar{\mathbb{D}^t}=\mathbb{S}\mathbb{D}^t\mathbb{S}^\mathsf{T}$,
where we will only consider terms up to second order in $\theta$ in
the following.  Furthermore we average over all possible orientations,
weighted by the probability density $p(\theta_i,t)\sim
\exp(-\theta_i^2/(4D^\theta_i t))$ due to rotational diffusion. As we
consider only small $\theta$ and $t\ll1$, it will now also be
sufficient to expand $p(\theta_i,t)$ up to second order in $t$. Hence,
the Gaussian probability distribution can be replaced by a uniform
distribution regarding correct integral boundaries.  The non-diagonal
entries are odd in $\theta_i$ and thus vanish. The average of the
diagonal entries is the central quantity for the calculation of the
mean square displacement:
\begin{align}
\mean{\bar{\mathbb{D}^t}_{ii}}_t&=\frac{1}{w_1w_2w_3}\int\limits_{-w_1/2}^{w_1/2}\hspace{-2mm}\diff\theta_1
\int\limits_{-w_2/2}^{w_2/2}\hspace{-2mm}\diff\theta_2
\int\limits_{-w_3/2}^{w_3/2}\hspace{-2mm}\diff\theta_3
\bar{\mathbb{D}^t}_{ii} 
\text{ ,}
\label{crossover3d_integrand}
\end{align}
where $w_i=\sqrt{24D^\theta_i t}$ is the width of the uniform
distribution interval. Only terms of even orders of $\theta_i$ will
contribute. That is, the integral in Eq.\ \ref{crossover3d_integrand}
will only lead to zeroth and second moments of the angular
distribution.  By the average action of the rotational diffusion,
$D_i$ transforms into an effective translational diffusion constant
$\mean{\bar{\mathbb{D}^t}_{ii}}_t$ over time $t$ in the fixed
laboratory coordinate space. Considering a vector $\mathbf
D=(D_1,D_2,D_3)$, the evolution of the effective, orientation averaged
vector of diffusion coefficients $\mean{\mathbf
  D}(t)=(\mean{\bar{\mathbb{D}^t}_{11}}_t,
\mean{\bar{\mathbb{D}^t}_{22}}_t,\mean{\bar{\mathbb{D}^t}_{33}}_t)$
can be expressed in a matrix form, not taking into account non-linear
terms in $t$:
\begin{align}
\mean{\mathbf D}(t)&=\mathbf D\cdot\mathbb{R}(t)+\mathcal O(t^2)\text{ ,}\\
\mathbb{R}_{ij}(t)&=
2t|\epsilon_{ijk}|D^\theta_k+\delta_{ij}(1+2t(D^\theta_i-\bar{D^\theta}))
\text{ ,}
\end{align}
where $\bar{D^\theta}=\sum_{k=1}^3D^\theta_k$. The principal axes of
effective motion are constant since the coupling terms between the
translational degrees of freedom vanish when averaging over all
possible orientations. Because rotational diffusion is independent of
time and orientation, $\mathbb R$ keeps its form for all times. We can
also apply $\mathbb R(\delta t)$ to some diffusion vector
$\mean{\mathbf D}(t)$. Thus, it is possible to evaluate the effective
change of the diffusion coefficients for large times $t$ in small
steps $\delta t=t/N$ with only making errors of $\mathcal O (N \delta
t^2)$:
\begin{align}
\mean{\mathbf D}(N\delta t)&=\mathbf D\cdot\left(\mathbb{R}(\delta
t)\right)^N+\mathcal O(N\delta t^2)\text{ .}
\label{crossover_infinit_error_definition}
\end{align}
In the limit of large $N$ the error in
\meqref{crossover_infinit_error_definition} vanishes: $\mathcal
O(Nt^2/N^2)=\mathcal O(t^2/N)\rightarrow0$.  This basically means that
we calculate $\mean{\mathbf D}(t)$ with infinite accuracy. We can now
evaluate the overall, orientation averaged mean square displacement by
integrating $\mean{\mathbf D}(t)$ with respect to $t$:
\begin{align}
\mean{\mathbf x(t)^2}&=
\begin{pmatrix}
2ft+(a d_2 -b d_3 ) g_- +c d_1  g_+ \\
2ft+(a d_1 -b d_2 ) g_- +c d_2  g_+ \\
2ft+(a d_3 -b d_1 ) g_- +c d_3  g_+ 
\end{pmatrix}\text{ ,}
\label{crossover_mainresult} 
\end{align}
with
$a=D^\theta_3-D^\theta_1$, 
$b=D^\theta_1-D^\theta_2$, 
$c=\sqrt{a^2+b^2+ab}$, 
$d_i=3(D_i-f)$, 
$f=\sum_{i=1}^3 D_i/3$, and
\begin{align}
 g_\pm&=
\dfrac{1-e^{-2(\bar{D^\theta}-c)t}}{6(\bar{D^\theta}-c)c}\pm\dfrac{1-e^{-2(\bar{D^\theta}+c)t}}{6(\bar{D^\theta}+c)c}
\label{crossover_finalgdef}
\text{ .} 
\end{align}
This result can be used to obtain effective translational diffusion
coefficients via $\mathbf D^\mathrm{eff}(\Delta t)=\mean{\mathbf
  x^2(\Delta t)}/2t$, compare Fig.~\ref{aniso_iso_compare}(b). Thus, the
crossover from anisotropic to isotropic diffusion in 3D occurs on two
time scales $T^c_{1/2}=1/(2\bar{D^\theta\pm c})$ which increasingly
deviate with increasing anisotropy.

The corresponding crossover times for ellipsoids are
$T^c_1=1/(4D^r_\parallel(\xi)+2D^r_\perp(\xi))$ and
$T^c_2=1/(6D^r_\perp(\xi))$. The rotational friction coefficients and
their implication for $T^c_{1/2}$ are shown in
Fig.~\ref{aniso_iso_compare}(c). The asymptotic behavior indicated in
the plot can be derived from the full solution of the friction
coefficients. For $\xi<0$ both $D^r_\parallel$ and $D^r_\perp$
approach a constant value. Therefore, regarding rotational diffusion
no significant differences are expected for oblates, which corresponds
well to the findings from Fig.~\ref{aniso_iso_compare}(a). In
contrast, rotational diffusion particularly around $\mathbf e_\perp$,
which governs $T^c_2$, is strongly decreased for large
$\xi$. Therefore, the relevant range of anisotropic diffusion grows
and the indication of the scaling in Fig.~\ref{aniso_iso_compare}(a)
shows that this again corresponds well to the simulation
data. Assuming this scaling to govern the impact of hydrodynamic
anisotropy, one might argue that the effect will be strongest for
$\xi\gg1$ and small $L$, i.e.\ large concentrations. However, if we
require $L>2\xi$, so that the particles fit into the simulation box,
encounter times are larger than $T^c_2$ for all values of $\xi$.
Thus our analytical calculation confirms that anisotropic
diffusion does not have a strong effect on protein-protein encounter
rates. This finding also validates computational schemes which
assume isotropic diffusion for efficient description of the
diffusion steps \cite{wolde_05_a}.

An interesting question that has not been addressed yet is whether the
effect of hydrodynamic anisotropy is stronger regarding re-encounter.
Dynamic dissociation and re-association can be easily realized in our
model for example by introducing a finite average lifetime for each
bond. Recently it has been shown with a similar Brownian dynamics
approach that productive protein-protein encounter is preceded by many
non-productive contacts \cite{schluttig_08}.  In a similar vein,
dissociation after binding is likely to be followed by additional
encounters. Because prolates have a lower overall rotational diffusion
coefficient, one expects that they are more likely to return to an
encounter. It might well be that for re-encounter, the aspect ratio
plays a more important role for accessibility of the binding site than
found here for protein diffusion to the first encounter. Similar
considerations might be valid for protein dynamics in the context
of protein clusters, where diffusion might be restricted by the
presence of other components.


\end{document}